 \documentclass[aip,pop,reprint]{revtex4-1}
 \usepackage{dcolumn}
 \usepackage{bm}
 \usepackage{amsmath}
 \usepackage{tabularx}

\newcommand{\erfc}{\textrm{erfc}}

\newcommand{\vc}{\mathbf}

\usepackage{graphicx}
\usepackage{color}

\begin{document}


\title{The incomplete plasma dispersion function: properties and application to waves in bounded plasmas}


\author{S.\ D.\ Baalrud}

\affiliation{Theoretical Division, Los Alamos National Laboratory, Los Alamos, New Mexico 87545}
\affiliation{Department of Physics and Astronomy, University of Iowa, Iowa City, IA 52242
}


\date{\today}

\begin{abstract}

The incomplete plasma dispersion function is a generalization of the plasma dispersion function in which the defining integral spans a semi-infinite, rather than infinite, domain. It is useful for describing the linear dielectric response and wave dispersion in non-Maxwellian plasmas when the distribution functions can be approximated as Maxwellian over finite, or semi-infinite, intervals in velocity phase-space.  A ubiquitous example is the depleted Maxwellian electron distribution found near boundary sheaths or double layers, where the passing interval can be modeled as Maxwellian with a lower temperature than the trapped interval. The depleted Maxwellian is used as an example to demonstrate the utility of using the incomplete plasma dispersion function for calculating modifications to wave dispersion relations. 

\end{abstract}




\maketitle


\section{Introduction\label{sec:intro}}

The incomplete plasma dispersion function, defined as
\begin{equation}
Z (\nu, w) \equiv \frac{1}{\sqrt{\pi}} \int_{\nu}^\infty dt \frac{e^{-t^2}}{t - w} ,  \label{eq:ipdf}
\end{equation}
for $\Im \lbrace w \rbrace > 0$ and as its analytic continuation for $\Im \lbrace w \rbrace \leq 0$, was introduced by Franklin.\cite{fran:71} It is a generalization of the plasma dispersion function,\cite{frie:61,fadd:61} which corresponds to the limit $\nu \rightarrow - \infty$: $Z(w) = Z(-\infty , w)$. Like the plasma dispersion function, it arises in the linear plasma dielectric function and corresponding wave dispersion relations. However, the plasma dispersion function describes Maxwellian distributions, whereas Eq.~(\ref{eq:ipdf}) describes non-Maxwellian distributions so long as they can be approximated as Maxwellian in finite, or semi-infinite, intervals of velocity phase-space. Each interval may have different characteristic densities, flow speeds and temperatures associated with them. 

\begin{figure}
\includegraphics{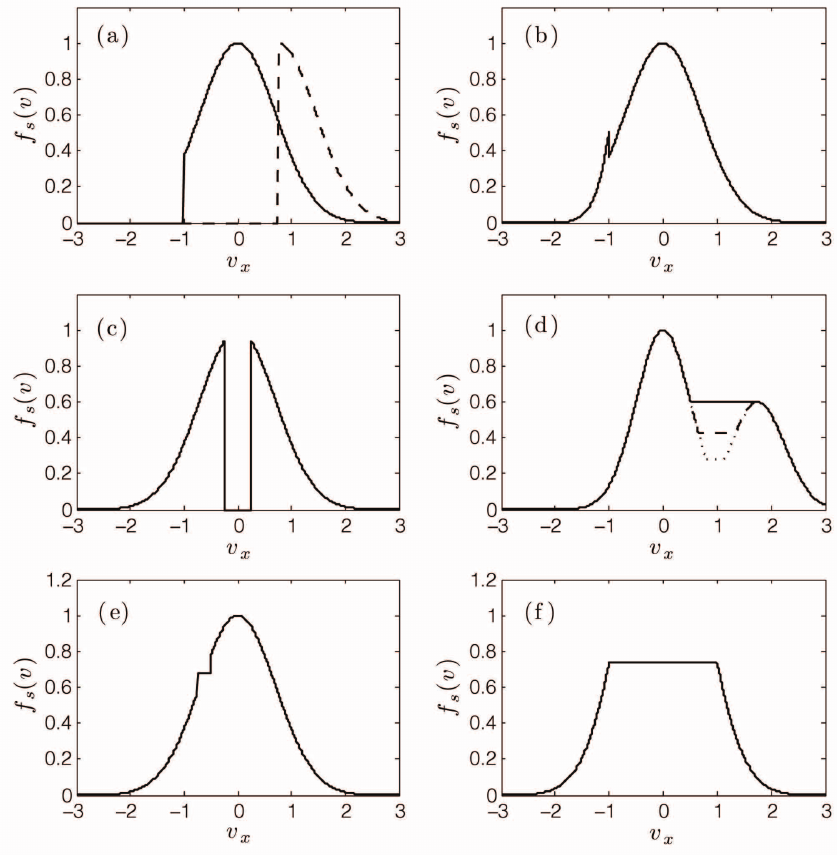}
\caption{Example distributions functions for which the incomplete plasma dispersion function will arise in the linear dielectric: (a) depleted Maxwellians with a flow-shift (dashed) and without (solid), (b) depleted Maxwellian with an additional tail population from secondary electron emission, (c) Maxwellian distribution with a hole at low energy, (d) plateau distributions, (e) Maxwellian with a narrow flattened interval, and (f) a flat-top distribution. }
\label{fg:ex_f}
\end{figure}

Situations where a plasma is far from equilibrium, but model distribution functions can be formulated in terms of piecewise Maxwellians are prevalent.\cite{cano:72,henr:72,treg:73,jaco:74,fran:75,treg:76,trot:77,phel:78,kuhn:81,gody:92,meig:06,boff:11,guo:12,dorf:09,baal:11b,baal:07,sydo:06,sydo:07,lieb:06,corr:08,scim:10,taka:07,baal:11a,kawa:09,aane:06,sten:08,baal:09,sten:11,hers:85,eged:05,eged:08,le:09,le:10,eged:12,scud:92} An early application was in the study of waves near the boundaries of Q-machine plasmas.\cite{fran:75,kuhn:81} There ions were modeled with a distribution consisting of only the high-energy tail of a Maxwellian. Electrons trapped by the confining ion sheaths were considered Maxwellian, whereas the passing interval was depleted in density. These distributions are depicted by the dashed and sold lines, respectively, in Fig.~\ref{fg:ex_f}a. 

This type of depleted electron distribution is commonly found in the presence of potential barriers, such as sheaths near material walls,\cite{gody:92,meig:06,boff:11,guo:12} or probes,\cite{dorf:09,baal:11b,baal:07} as well as double layers which provide an electrostatic barrier between plasmas with differing properties (density, temperature, etc.).\cite{lieb:06,corr:08,scim:10,taka:07,baal:11a,kawa:09,aane:06,sten:08,baal:09,sten:11,hers:85} Potential barriers create a trapped-passing boundary for one or more species. At the edge of an absorbing wall sheath, for example, the passing interval of the electron distribution will be empty. This depleted region can significantly alter the plasma dielectric response. For example, it has been shown to reduce the threshold for temperature-anisotropy-driven whistler instabilities near the boundary of magnetized plasmas.\cite{guo:12} Scattering can fill in the depleted region such that the density of the passing interval increases away from the potential barrier. Also, in some applications, secondary electron emission can generate an additional population of tail electrons,\cite{sydo:06,sydo:07} see Fig.~\ref{fg:ex_f}b, which may also be modeled using Eq.~(\ref{eq:ipdf}). 

Other examples include electron sheaths, double sheaths and double layers that arise when a boundary is biased more positive than the plasma potential.\cite{sten:08,baal:09,sten:11} In these situations the electron distribution can often be modeled as a depleted half-Maxwellian with a flow shift; see Fig.~\ref{fg:ex_f}a. Phelps and Allen studied waves in the presence of a double sheath with electron emitting boundaries.\cite{phel:78} They modeled the electron distribution as a background Maxwellian with an additional flowing half-Maxwellian component generated by wall emission. By modeling the electrons in this manner, they were able to accurately predict measured wave dispersion relations for high frequency waves. They noted that ions in these configurations can resemble Maxwellian distributions with a hole at low energies, as shown in Fig.~\ref{fg:ex_f}c. 

The incomplete plasma dispersion function may also be useful for studying waves after non-linear process have modified a velocity distribution function, such as flattening of a region of velocity-space due to a driven wave or instability,\cite{vale:12} as shown in Fig.~\ref{fg:ex_f}e. Another common example is the plateau distribution following the quasilinear evolution of a bump-on-tail distribution, as depicted in Fig.~\ref{fg:ex_f}d. These have been measured for ions downstream of current-free double layers.\cite{corr:08,scim:10} Piecewise Maxwellian distributions have also been used to model electrons upstream of current-free double layers,\cite{lieb:06,taka:07,baal:11a,aane:06} as well as double layers that form in electronegative plasmas.\cite{kawa:09}

The incomplete plasma dispersion function may also find use in magnetic reconnection research. A recent theory proposes that large scale parallel electric fields can both accelerate electrons and create a trapped-passing barrier that affects the electron distribution.\cite{eged:05,eged:08,le:09,le:10,eged:12} In this model, the parallel electron distribution is Maxwellian in the tails and constant in the intermediate region, as shown in Fig.~\ref{fg:ex_f}f. Notably, Eq.~(\ref{eq:ipdf}) will arise in the dispersion relation of collisionless tearing modes, which can be a trigger for fast reconnection in magnetospheric plasmas. 

The purpose of this paper is to provide several properties of the incomplete plasma dispersion function that are useful for applying it to the linear dielectric response and wave dispersion relations in piecewise Maxwellian plasmas. These properties are discussed in Sec.~\ref{sec:prop}. In Secs.~\ref{sec:vsd} and \ref{sec:pvsd}, the utility of this function is demonstrated by applying it to the application of electrostatic waves near absorbing boundaries of unmagnetized plasmas. In Sec.~\ref{sec:vsd}, the depleted region of the electron distribution is taken to be completely devoid of particles, whereas in Sec.~\ref{sec:pvsd} the depleted region is modeled as a colder Maxwellian (returning the results of Sec.~\ref{sec:vsd} in the limit of zero tail temperature). Waves in plasmas with gently depleted tails,\cite{gune:01} as well as Lorentzian (or kappa) distributions\cite{summ:91,hell:02} have also been studied previously.

Ion-acoustic and Langmuir wave dispersion relations are shown to be significantly modified when the trapped-passing boundary affects the bulk distribution (i.e., it is near the thermal speed), and the temperature characterizing the passing population is low enough. For instance, one significant modification to Langmuir waves is the absence of Landau damping for modes with phase velocity beyond the trapped-passing boundary because of the lack of resonant electrons. Similar undamped (or weakly damped) waves have previously been studied in the high frequency limit near a floating probe using a total depletion model. The modeled wave properties were also confirmed using a probe and spectrum analyzer.\cite{treg:76} Similarly, the dispersion relations developed in Secs.~\ref{sec:vsd} and \ref{sec:pvsd} should be accessible to experimental measurement with the classical probe techniques. In addition, modern laser-induced fluorescence techniques may be capable of measuring both the dispersion properties of these waves, as well as how they may influence the distribution functions in velocity-space.\cite{bach:98,skiff:98,skiff:02}


\section{Properties\label{sec:prop}} 

\subsection{Differential representation}

The incomplete plasma dispersion function, Eq.~(\ref{eq:ipdf}), can also be represented in the differential form  
\begin{equation}
\frac{dZ}{dw} + 2 w Z = \frac{1}{\sqrt{\pi}} \frac{e^{-\nu^2}}{\nu - w} - \erfc(\nu) . \label{eq:ode}
\end{equation}
Equation~(\ref{eq:ode}) can be used to write higher order derivatives in terms of lower-order derivatives
\begin{equation}
Z^{(n)} = \frac{(n-1)!\, e^{-\nu^2}}{\sqrt{\pi} (\nu -w )^n} - 2 \bigl[ (n-1) Z^{(n-2)} + w Z^{(n-1)} \bigr] 
\end{equation}
for $n \geq 2$. Here $Z^{(n)} \equiv d^n Z/dw^n$ denotes the $n^\textrm{th}$ derivative of $Z$ with respect to $w$. 

\subsection{Asymptotic expansions}

Asymptotic expansions can be obtained by applying the Plemelj formula 
\begin{equation}
Z(\nu,w) \simeq i \sqrt{\pi} H(w-\nu) e^{-w^2} + \frac{1}{\sqrt{\pi}} \mathcal{P} \int_\nu^\infty dt \frac{e^{-t^2}}{t-w} , \label{eq:plemelj}
\end{equation}
in which $H$ is the Heaviside step function, then using a large or small argument Taylor expansion of the integrand. For $|w| \ll 1$, this yields 
\begin{equation} 
Z(\nu, w) = i \sqrt{\pi} H(w - \nu) e^{-w^2} + \sum_{n=0}^\infty a_n w^{n} \label{eq:sarg}
\end{equation}
in which 
\begin{equation}
a_n = \frac{\Gamma (-n/2, \nu^2) + 2 \alpha (n+1 , \nu) \gamma(-n/2 , \nu^2)}{2 \sqrt{\pi}}  . \label{eq:an}
\end{equation}
Here
\begin{equation}
\alpha (n,\nu) \equiv [1+(-1)^{n}] (1 - \nu/|\nu| )/4
\end{equation}
is a parameter that is unity if $\nu < 0$ and $n$ is even, but is zero otherwise, $\Gamma (n,x)$ is the upper incomplete gamma function, and $\gamma ( n,x)$ is the lower incomplete gamma function.\cite{abra:65} The second term in Eq.~(\ref{eq:an}) is taken to vanish for even $n$ (due to $\alpha =0$), although $\gamma(-n/2, \nu^2)$ diverges for even $n$. The first four terms of the coefficient $a_n$ are
\begin{subequations}
\begin{eqnarray}
a_o &=& E_1 (\nu^2)/(2\sqrt{\pi}) , \\
a_1 &=& e^{-\nu^2}/(\nu \sqrt{\pi}) - \erfc (\nu) , \\ 
a_2 &=& e^{-\nu^2}/(2 \sqrt{\pi} \nu^2) - E_1(\nu^2)/(2 \sqrt{\pi}) ,\\
a_3 &=& \frac{1-2\nu^2}{3\sqrt{\pi} \nu^3} e^{-\nu^2} + \frac{2}{3} \erfc (\nu) ,
\end{eqnarray}
\end{subequations}
where $\erfc$ is the complimentary error function and $E_1 (\nu^2) = \Gamma (0, \nu^2)$ is the exponential integral.\cite{abra:65} 

Applying the same procedure in the asymptotic limit $|w| \gg 1$ yields
\begin{equation}
Z(\nu, w) \sim i \sigma \sqrt{\pi} H(w-\nu) e^{-w^2} - \sum_{n=1}^\infty b_n/w^n \label{eq:larg}
\end{equation}
in which
\begin{equation}
b_n = \frac{\Gamma (n/2, \nu^2) + 2 \alpha (n+1, \nu) \gamma (n/2, \nu^2)}{2 \sqrt{\pi}} , \label{eq:bn}
\end{equation}
and where 
\begin{eqnarray}
\sigma \equiv
\left\lbrace \begin{array}{ll}
0 ,  & \Im \lbrace w \rbrace > 0 , \\
1 ,  & \Im \lbrace w \rbrace = 0 , \\
2,  &  \Im \lbrace w \rbrace < 0 .
\end{array} \right. 
\end{eqnarray}
Note that $b_n = a_{-n}$. The first five terms of the coefficient $b_n$ are
\begin{subequations}
\begin{eqnarray}
b_1 &=& \erfc(\nu)/2 , \\ 
b_2 &=& e^{-\nu^2}/(2 \sqrt{\pi}) , \\ 
b_3 &=& \nu e^{-\nu^2}/(2 \sqrt{\pi}) + \erfc (\nu)/4 ,\\ 
b_4 &=& (1 + \nu^2)e^{-\nu^2}/(2 \sqrt{\pi}) , \\
b_5 &=& \frac{\nu^3 + 3\nu /2}{2 \sqrt{\pi}} e^{-\nu^2} + \frac{3}{8} \erfc( \nu ) .
\end{eqnarray}
\end{subequations}

\subsection{Generalized functions\label{sec:gf}} 

Generalized functions of the form
\begin{equation}
Z_n (\nu, w) \equiv \frac{1}{\sqrt{\pi}} \int_\nu^\infty dt \frac{t^n e^{-t^2}}{t-w} , \ \ n \geq 0, \ \Im \lbrace w \rbrace > 0 \label{eq:zgen}
\end{equation}
can be related to derivatives of $Z (\nu, w)$ via the generating function
\begin{align}
Z_n &= \frac{1}{2^n} \sum_{l=0}^n (-1)^l d_l(n) \biggl[ Z^{(l)}  \label{eq:zngen}  \\ \nonumber
& - \frac{1}{\sqrt{\pi}} \sum_{m=0}^{l-1} \frac{(l-m-1)!\, (-1)^m H_m(\nu) e^{-\nu^2}}{(\nu - w)^{l-m}} \biggr] ,
\end{align}
where $H_m$ denotes the $m^\textrm{th}$ Hermite polynomial, and $d_l(n)$ denote the coefficients satisfying $t^n = 2^{-n} \sum_{l=0}^n d_l (n) H_l (t)$; see Tbl.~\ref{tb:dlm} (from Tbl.~22.12 of Ref.~\onlinecite{abra:65}).  A derivation of Eq.~(\ref{eq:zngen}) is provided in appendix \ref{sec:zngen}. The first few terms of Eq.~(\ref{eq:zngen}) are
\begin{subequations}
\begin{eqnarray}
Z_0 (\nu,w) &=& Z , \\
Z_1 (\nu,w) &=& - \frac{Z^\prime}{2} + \frac{e^{-\nu^2}}{2 \sqrt{\pi} (\nu-w)} = \frac{\erfc (\nu)}{2} + w Z ,  \\
Z_2 (\nu,w) &=& \frac{1}{4} \biggl[ 2 Z + Z^{\prime \prime} + \frac{2 \nu (\nu - w) - 1}{\sqrt{\pi} (\nu - w)^2} e^{-\nu^2} \biggr] \\ \nonumber
 &=& w [ \erfc (\nu)/2 + w Z ] + e^{-\nu^2}/(2 \sqrt{\pi}) .
\end{eqnarray}
\end{subequations}

\begin{table}
\caption{Values for the coefficients $d_l (n)$ and $c_n(l)$. The top row gives values of $n$, and the left column values of $l$ [e.g., $d_3(5) = 20]$. The Hermite polynomials are generated from $H_l = \sum_{n=0}^l c_n (l) x^n$ [e.g., $c_2(4)=-48$]. For $l=n$ the left value corresponds to $c_n(l)$ and the right value to $d_l(n)$. }
\begin{center}
\begin{tabular}{r | c | c | c | c | c | c | c}
\hline \hline
\multicolumn{8}{c}{$n$ values} \\
\hline
$l$ values   &  0    & 1   &  2     & 3     &   4   & 5  & 6 \\ \hline
0     & 1\, \, 1     & 0     &  2   & 0     & 12  & 0  & 120 \\ \hline
1     & 0         & 2\, \, 1 &  0   & 6     &   0   & 60 &  0 \\  \hline 
2     &  -2       & 0    & 4\, \, 1 & 0     &   12 & 0   & 180 \\ \hline
3     &  0        &-12  & 0    & 8\, \, 1 &   0    & 20 &    0 \\ \hline
4     &  12      &  0  & -48  &  0  &16\, \, 1 & 0   &  30  \\ \hline
5     &  0        &120 & 0   &-160&   0     & 32\, \, 1   &  0  \\ \hline
6     &  -120  &   0  &720 &  0   & -480 &   0    &64\, \, 1  \\ 
\hline \hline
\end{tabular}
\end{center}
\label{tb:dlm}
\end{table}

\subsection{Numerical evaluation\label{sec:num}}

The incomplete plasma dispersion function can be evaluated using similar techniques used to evaluate the plasma dispersion function. When the pole at $t=w$ falls outside the integration interval ($w < \nu$), Eq.~(\ref{eq:ipdf}) can computed by direct integration. When the pole at $t=w$ falls within the integrand (i.e., $\nu < w < \infty$), a similar integral spanning $-\infty$ to $\nu$ can be computed by direct integration, and the result subtracted from the plasma dispersion function to yield the incomplete plasma dispersion function
\begin{equation}
Z(\nu ,w ) = Z(w) - \frac{1}{\sqrt{\pi}} \int_{-\infty}^\nu dt \frac{e^{-t^2}}{t - w}  .
\end{equation}
Efficient methods, such as continued fraction expansions (see Sec.~\ref{sec:cfrac}), have been developed for evaluating $Z(w)$. 

\begin{figure}
\includegraphics{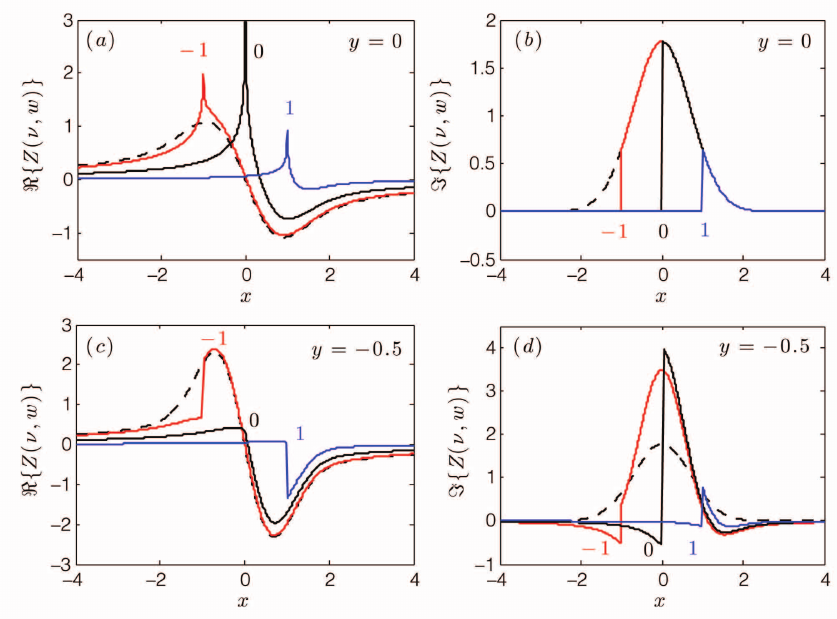}
\caption{Real and imaginary components of $Z (\nu, w)$ in which $w = x + iy$. Solid lines correspond to various values of $\nu$: $\nu = -1$ (red), $\nu = 0$ (black) and $\nu = 1$ (blue). The dashed line shows the complete plasma dispersion function $Z(-\infty , w)$. (a) and (b) take $y = 0$ and (c) and (d) take $y = -0.5$.  }
\label{fg:icpdf}
\end{figure}

\begin{figure*}
\includegraphics{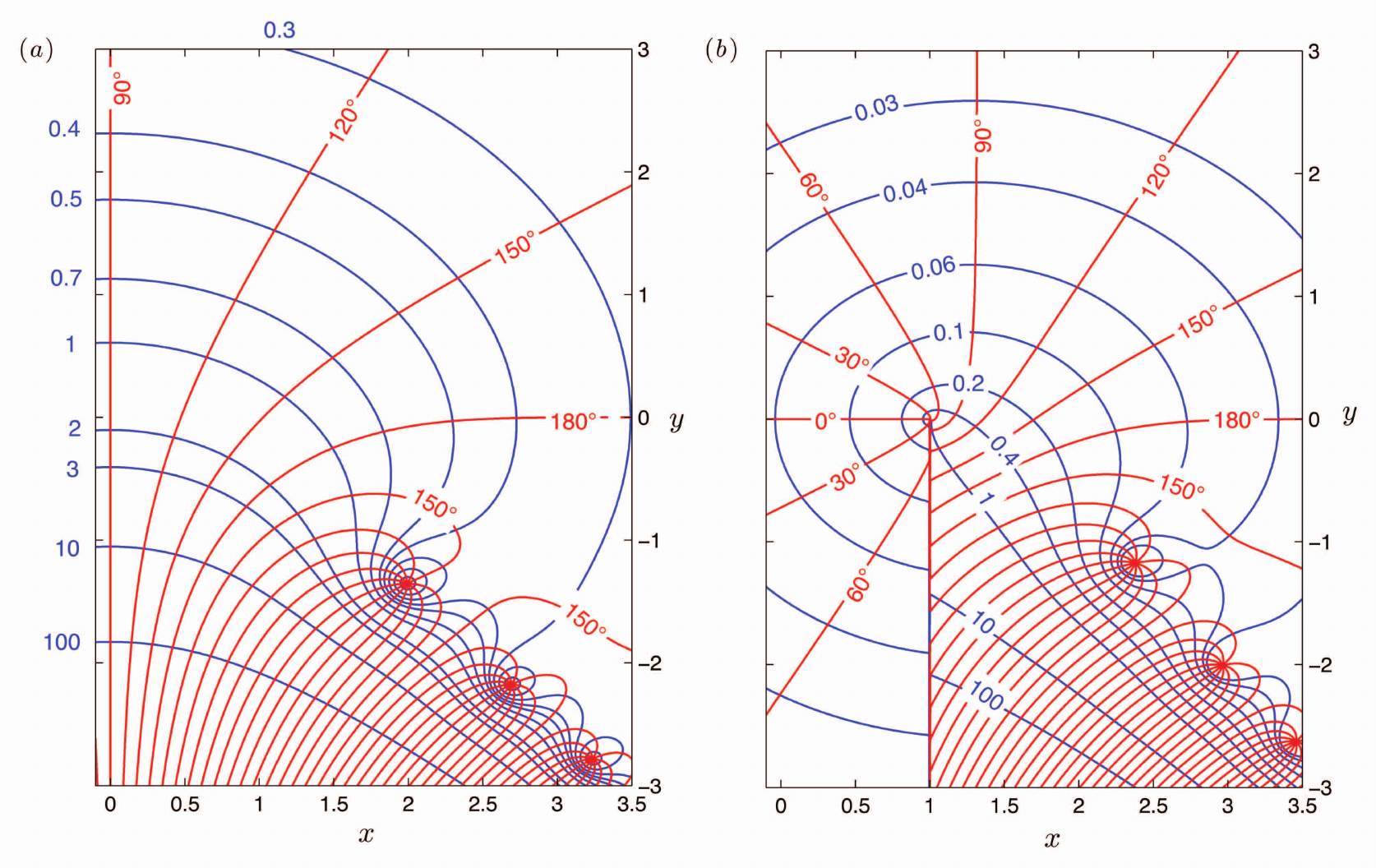}
\caption{Altitude plot of (a) $Z(-\infty, w)$ and (b) $Z(1, w)$ where $w = x + iy$.  }
\label{fg:alt}
\end{figure*}

Figure~\ref{fg:icpdf} shows plots of the real and imaginary parts of the incomplete plasma dispersion function as a function of the complex argument $w = x + i y$ for three values of the cutoff parameter $\nu = -1, 0$ and $1$. The complete plasma dispersion function is also shown as a dashed line for reference. For cuts along the real line ($y=0$) the real part of $Z(\nu, w)$ has a pole at $x = \nu$, as shown in panel (a). As $w \rightarrow \nu$, the function diverges logarithmically. This can be seen by considering the principal value term of Eq.~(\ref{eq:plemelj}), applying the substitution $s = t -w$ and expanding the exponential term for $s \ll w$ in the dominant interval near $s =0$. The leading term shows that the integral scales as $-\exp (-w^2) \ln (| \nu - w|)$ as $w \rightarrow \nu$.  The imaginary part is equal to $\sqrt{\pi} H(x - \nu) \exp (-x^2)$ for $y=0$, as show in panel (b). For cuts along the line $y=-0.5$, the real part closely follows the complete plasma dispersion function for $x>\nu$, and no singularities arise, as shown in panel (c). The imaginary part is again discontinuous at $x=\nu$, but has a more complicated behavior than the simple Gaussian shape found along the real line, as shown in panel (d). 

Figure~\ref{fg:alt} shows an altitude plot of the complete plasma dispersion function and incomplete plasma dispersion function for $\nu = 1$ in panels (a) and (b) respectively. The plot is obtained from the representation $Z = |Z| e^{i\theta}$ where blue lines are lines of constant $|Z|$ and red lines are lines of constant $\theta$. In the incomplete case, a branch cut arises at $x = \nu$ creating a discontinuous boundary for constant $|Z|$ lines (see Refs.~\onlinecite{kent:74,cano:74,godf:76} for a discussion). The intersection of this and the real line locates a singular point in $Z(\nu, w)$ that is not present in the complete case. 

\subsection{Two-pole approximation}

Because the plasma dispersion function is difficult to deal with analytically, it is often useful to have an approximate algebraic expression. A method that has been shown to accurately capture both the real and imaginary parts of the conventional plasma dispersion function is the two-pole approximation.\cite{frie:68,mart:79} This method can be extended to the incomplete plasma dispersion function. The approach is to first write a Pad\'{e} approximate of the form
\begin{equation}
\tilde{Z}(\nu, w) = \frac{p_o + p_1 w}{1 + q_1 w + q_2 w^2} \label{eq:pade}
\end{equation}
in which the coefficients are determined by matching the large and small argument asymptotic expansions of Eq.~(\ref{eq:pade}) with those of Eq.~(\ref{eq:ipdf}). Using Eq.~(\ref{eq:larg}), the large argument limit gives $p_1/q_2 = b_1$. Using Eq.~(\ref{eq:sarg}), the small argument expansion of $Z(\nu, w)$ is $Z \simeq \sum_{n=0}^\infty c_n w^n$, where
\begin{equation}
c_o = i \sqrt{\pi} H(x - \nu) + \frac{E_1 (\nu^2)}{2 \sqrt{\pi}} \ \ , \ \ c_1 = \frac{e^{-\nu^2}}{\nu \sqrt{\pi}} - \erfc (\nu) 
\end{equation}
and
\begin{equation}
c_2 = -i\sqrt{\pi} H(x - \nu) + \frac{e^{-\nu^2}}{2 \sqrt{\pi} \nu^2} - \frac{E_1 (\nu^2)}{2 \sqrt{\pi}} .
\end{equation}
Matching these with the small argument expansion of Eq.~(\ref{eq:pade}), the $p$ coefficients are
\begin{equation}
p_o = c_o \ \ , \ \ p_1 = \frac{b_1 (c_1^2 - c_o c_2)}{c_o^2 + c_1 b_1} 
\end{equation}
and the $q$ coefficients are
\begin{equation}
q_1 = - \frac{c_o c_1 + c_2 b_1}{c_o^2 + c_1 b_1} \ \ , \ \ q_2 = \frac{c_1^2 - c_o c_2}{c_o^2 + c_1 b_1} .
\end{equation}

\begin{figure}
\includegraphics{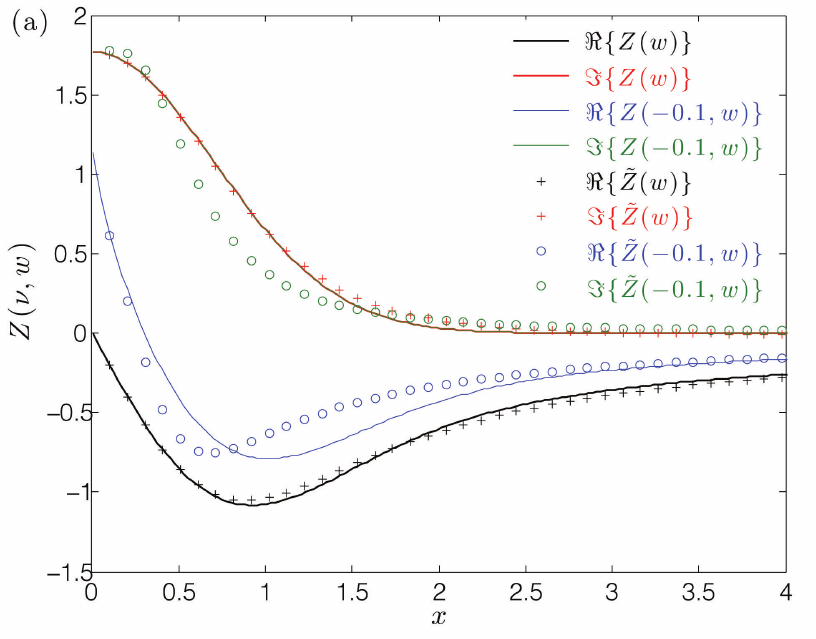}
\includegraphics{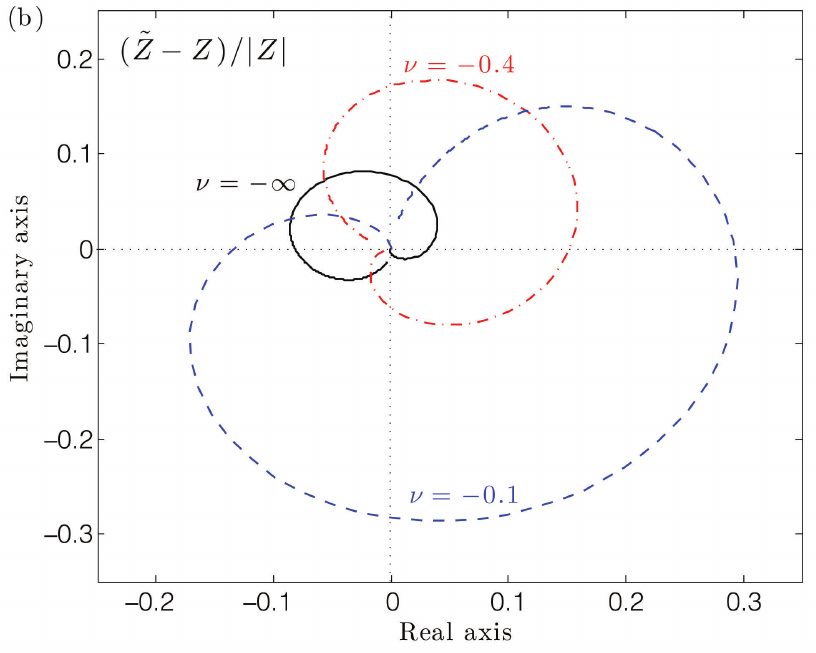}
\caption{(a) Comparison of the numerically evaluated $Z$ and the two-pole approximation $\tilde{Z}$ from Eq.~(\ref{eq:2pole}) for the complete case ($\nu = -\infty$) and an incomplete case ($\nu = -0.1$) taking a real argument $w = x + 0 i$. The lines representing numerical evaluation of the imaginary components lie on top of one another. (b) A measure of the relative error of the two-pole approximation [$(\tilde{Z} - Z)/|Z|$] for three values of the cutoff parameter $\nu = -\infty,\ -0.4$ and $-0.1$. Again, the argument is real and $x:\ 0 \rightarrow 20$. }
\label{fg:2pole}
\end{figure}

Next, a two-pole form of Eq.~(\ref{eq:pade}), written as
\begin{equation}
\tilde{Z} (\nu , w) = \frac{A}{c_+ - w} + \frac{B}{c_- - w} , \label{eq:2pole}
\end{equation} 
can be obtained from its partial fraction decomposition. This provides
\begin{equation}
c_{\pm} = - \frac{q_1}{2 q_2} \pm \frac{\sqrt{q_1^2 - 4q_2}}{2 q_2} 
\end{equation}
and
\begin{equation}
A = - \frac{p_o + p_1 c_+}{\sqrt{q_1^2 - 4 q_2}} \ \ , \ \ B = \frac{p_o + p_1 c_-}{\sqrt{q_1^2 - 4q_2}}  .
\end{equation}
Equation~(\ref{eq:2pole}) is a two-pole approximation of the incomplete plasma dispersion function. In the complete limit ($\nu \rightarrow -\infty$), the coefficients simplify considerably: $A = 0.5 -1.289 i$, $B=A^*$ and $c_\pm = \mp 0.514 - 1.032 i$.\cite{mart:79}

A comparison of the two-pole approximation from Eq.~(\ref{eq:2pole}) and numerical solutions of Eq.~(\ref{eq:ipdf}) are shown in Fig.~\ref{fg:2pole}a. Here, the complete case ($\nu = -\infty$) and an incomplete case $(\nu = -0.1)$ are shown and the argument is taken to be real $w = x$. This Pad\'{e}-type approximation captures both the large and small argument limits, and smoothly connects them. However, the incomplete plasma dispersion function has a pole on the real line at $w = \nu$ (see Fig.~\ref{fg:icpdf}a), which this approach does not capture. This leads to significant errors in the vicinity of the pole. This technique has a limited range of validity for the incomplete plasma dispersion function, and one must use it cautiously. On the other hand, as Fig.~\ref{fg:2pole} shows, it can still be useful when the argument is far from the pole. A common situation where this approximation may be useful is when $\nu < 0$, but one is interested in waves for which $w_R > 0$. 

Figure \ref{fg:2pole}b shows a measure of the relative error $(\tilde{Z} - Z)/|Z|$ between numerical solutions and two-pole approximations, again taking a real argument which here varies from 0 to 20. The error grows as the cutoff parameter is increased from approximately $10\%$ maximum in the complete case to approximately $30\%$ for $\nu = -0.1$. 

\subsection{Continued fraction expansion\label{sec:cfrac}}

The two-pole approximation of the previous section was based on a Pad\'{e} approximate written in a rational-fraction form
\begin{equation}
Z (\nu, w) \simeq P_M^N (w) = \sum_{n=0}^N A_n w^n \biggl/ \sum_{n=0}^M B_n w^n \label{eq:rational}
\end{equation}
where $N=1$ and $M=2$. This approach can be extended to higher order to obtain a convergent approximation. Equivalently, the Pad\'{e} representation can be written as a continued fraction. This is a common method used to numerically evaluate the plasma dispersion function. In this section, a continued fraction expansion of Eq.~(\ref{eq:ipdf}) and algorithm for evaluating it are calculated following the procedure presented in Ref.~\onlinecite{mcca:76} (for an alternative method, see Ref.~\onlinecite{pera:84}).\cite{supp} The result is also used to calculate a rational fraction representation.

McCabe and Murphy\cite{mcca:76} have developed an algorithm for calculating a continued fraction expansion of a function $f(z)$ corresponding to power series expansions at two points. The expansions are assumed to have the form
\begin{equation}
f(z) = \sum_{n=0}^\infty \bar{a}_n w^n  \label{eq:power}
\end{equation}
and 
\begin{equation}
f(z) = - \sum_{n=1}^\infty \bar{b}_n/w^n   \label{eq:asympt}
\end{equation}
with $\bar{a}_0, \bar{a}_1 \neq 0$. For our application, Eq.~(\ref{eq:power}) will correspond to the small argument expansion and Eq.~(\ref{eq:asympt}) to the large argument expansion of Eq.~(\ref{eq:ipdf}). 

The continued fractions are of the form
\begin{equation}
F (w) = \cfrac{\bar{a}_0}{1 + d_1^{(0)} w + \cfrac{n_2^{(0)} w}{1 + d_2^{(0)} w + \cfrac{n_3^{(0)} w}{1 + d_3^{(0)} w + \ldots  } }}  . \label{eq:cont_frac}
\end{equation}
The coefficients $n_i^{(0)}$ and $d_i^{(0)}$ can be calculated from the recurrence relations\cite{mcca:76}
\begin{subequations}
\begin{eqnarray}
n_{i+1}^{(r)} + d_i^{(r)} &=& n_i^{(r+1)} + d_i^{(r+1)} , \label{eq:nd1} \\ 
n_{i+1}^{(r)} d_{i+1}^{(r+1)} &=& n_{i+1}^{(r+1)} d_{i}^{(r)} \label{eq:nd2}
\end{eqnarray}
\end{subequations}
for $i=1,2,3,\ldots$ and $r=0, \pm 1, \pm 2, \ldots$ using the initial values $d_1^{(0)} = - \bar{a}_0/b_1$, $d_1^{(r)} = -\bar{a}_r/\bar{a}_{r-1}$ for $r \geq 1$, $d_1^{(r)} = - \bar{b}_{|r|}/ \bar{b}_{|r -1|}$ for $r \leq -1$ and $n_1^{(r)} = 0$ for all $r$. Solving Eqs.~(\ref{eq:nd1}) and (\ref{eq:nd2}) provides an array of terms, called the n-d array. Once these are determined, the continued fraction of Eq.~(\ref{eq:cont_frac}) can be evaluated to order $N$. 

\begin{figure}
\includegraphics{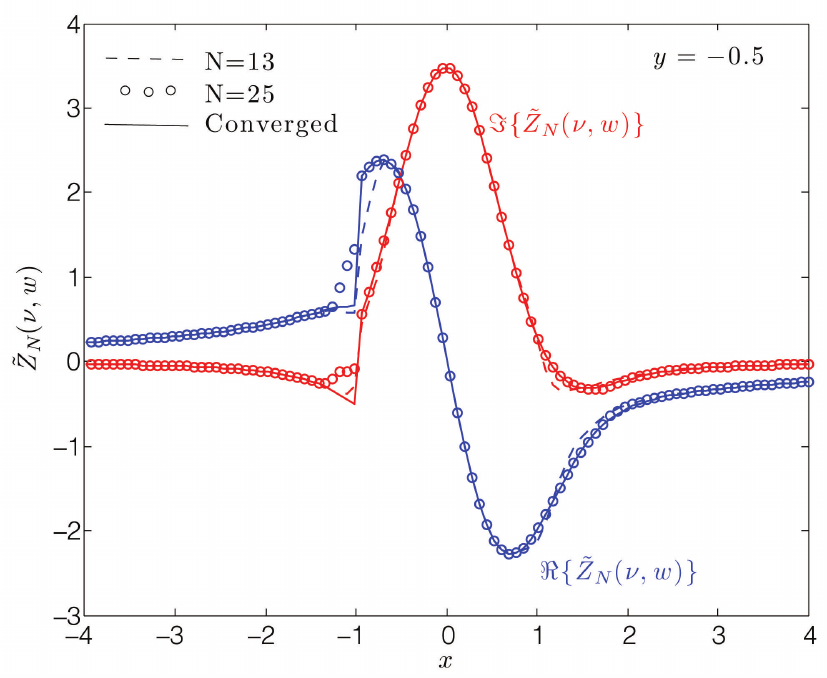}
\caption{Real (blue) and imaginary (red) components of the $N^\textrm{th}$ convergent of the continued fraction expansion of the incomplete plasma dispersion function $\tilde{Z}_N (\nu, w)$ for $\nu = -1$ and $w = x - 0.5 i$. Solid lines show a converged solution obtained using the method of Sec.~\ref{sec:num}. }
\label{fg:cfrac}
\end{figure}

McCabe and Murphy developed a convenient method to determine the convergents $F_1, F_2, \ldots , F_N$, which also provides rational Pad\'{e} approximates of the form of Eq.~(\ref{eq:rational}). The $N^\textrm{th}$ convergent ($F_N$), corresponding to the continued fraction of Eq.~(\ref{eq:cont_frac}) keeping terms $i=1,2,\ldots,N$, can be written in the rational fraction form
\begin{equation}
F_N (w) = R_N (w)/S_N(w) .
\end{equation}
Here $R_N$ and $S_N$ can be determined from the recurrence relations\cite{mcca:76}
\begin{subequations}
\begin{eqnarray}
R_{N+1} &=& (1 + d_{N+1}^{(0)} w ) R_N + n_{N+1}^{(0)} w R_{N-1} , \\
S_{N+1} &=& (1 +  d_{N+1}^{(0)} w) S_N + n_{N+1}^{(0)} w S_{N-1} 
\end{eqnarray}
\end{subequations}
for $N \geq 1$ and with the initial conditions $R_0 = 0$, $R_1 = \bar{a}_0$, $S_0 = 1$, and $S_1 = 1 + d_1^{(0)} w$. 

Because the asymptotic expansions of $Z(\nu, w)$ depend on $H(w_R - \nu)$, it is convenient to use the representation
\begin{equation}
Z (\nu, w) = i \sqrt{\pi} H(w_R - \nu) e^{-w^2} + f (\nu, w)  \label{eq:zcont}
\end{equation}
and approximate $f(\nu, w)$ with a continued fraction expansion; $f(\nu, w) \simeq F_N (\nu, w)$. In this case, the coefficients of Eqs.~(\ref{eq:power}) and (\ref{eq:asympt}) are simply the large and small argument expansions from Eqs.~(\ref{eq:an}) and (\ref{eq:bn}) (i.e., $\bar{a}_n = a_n$ and $\bar{b}_n = b_n$). Alternatively, the exponential term in Eq.~(\ref{eq:zcont}) could be included in the expansion coefficients, but then the coefficients become functions of $w_R$ through the $H (w_R -\nu)$ term. 

Figure~\ref{fg:cfrac} shows convergents of Eq.~(\ref{eq:zcont}) for $N=13$ and $25$ obtained from the continued fraction method. Here $\nu = -1$ and $y=-0.5$. Also shown is a converged numerical solution obtained using the method of Sec.~\ref{sec:num}. The continued fraction solution converges quickly over much of the domain, with the exception of points in the vicinity of large gradients. Since it is very difficult to capture regions with steep gradients using Pad\'{e} approximates, we have found the method of Sec.~\ref{sec:num} to provide a more efficient means for numerical evaluation. The advantage arises from splitting the integral into a smooth part (the complete plasma dispersion function), which can be calculated efficiently with continued fractions, and a part with steep gradients, which can be evaluated by direct integration.

\section{Waves in the presence of a depleted electron distribution\label{sec:vsd}} 

\begin{figure}
\begin{center}
\includegraphics{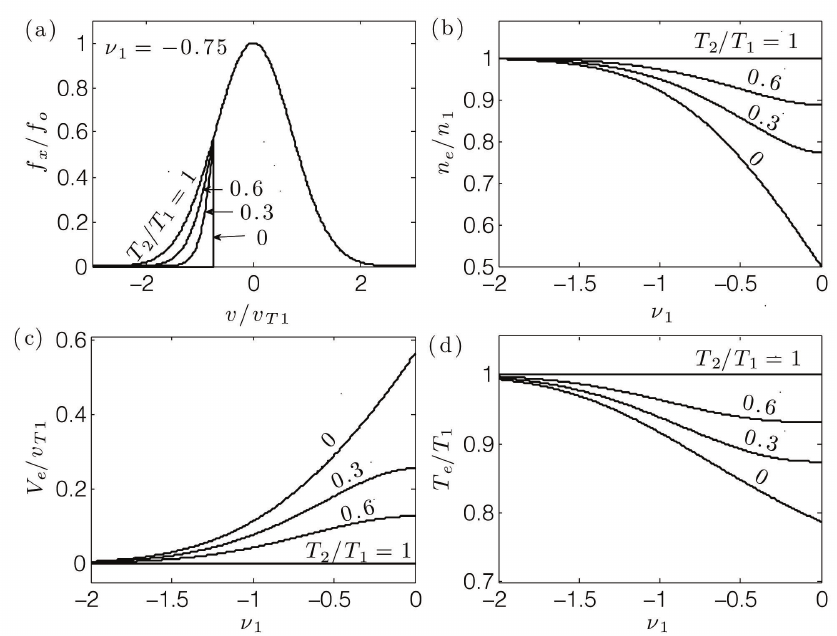}
\caption{(a) Depleted Maxwellian distribution for $\nu_1 = v_c/v_{T1} = -0.75$ and four values of $T_2/T_1$: 0, 0.3, 0.6 and 1. Also shown are the corresponding fluid moment parameters: (b) density, (c) flow speed, and (d) temperature. The case of total depletion ($T_2/T_1 = 0$) is used in Sec.~\ref{sec:vsd}. }
\label{fg:moments}
\end{center}
\end{figure}

Next, Eq.~(\ref{eq:ipdf}) is applied to calculate the dispersion relation of linear electrostatic waves near an absorbing boundary. The electron distribution in this scenario will have a trapped-passing boundary corresponding to the local electrostatic potential; see Fig.~\ref{fg:moments}a. Because the boundary is completely absorbing, the passing interval will nominally be empty. Collisions act to fill in the depleted interval, which populates as a function of distance from the boundary. In this section, we take the passing interval to be completely devoid of particles.  In Sec.~\ref{sec:pvsd} we will generalize this to include a passing population with an independent temperature.

The dispersion relation of linear electrostatic waves can be computed from the roots of the dielectric function
\begin{equation}
\hat{\varepsilon} = 1 + \sum_s \frac{4\pi q_s^2}{k^2 m_s} \int d^3v \frac{\vc{k} \cdot \nabla_\vc{v} f_s}{\omega - \vc{k} \cdot \vc{v}} \label{eq:dielec} .
\end{equation}
In this section, we take the electron distribution function to be
\begin{equation}
f_e = \frac{n_{1} H(v_x - v_c)}{\pi^{3/2} v_{T1}^3} e^{-v^2/v_{T1}^2} ; \label{eq:ftrun}
\end{equation}
see Fig.~\ref{fg:moments}a. In Eq.~(\ref{eq:ftrun}), $n_1$ and $T_1$ are parameters associated with the Maxwellian region of velocity space. The subscript $1$ is used to denote this interval. In this case, the passing interval (interval $2$) is empty so we need not consider it. We use this notation for consistency with the next section. The parameters $n_1$ and $T_1$ can be related to the density, flow velocity, and temperature of the distribution through the moment definitions: $n_s = \int d^3v\, f_s$, $\vc{V}_s = \int d^3v\, \vc{v} f_s/n_s$ and $T_s = \int d^3v\, m_s (\vc{v} - \vc{V}_s)^2 f_s/(3n_s)$. Inserting Eq.~(\ref{eq:ftrun}) provides
\begin{equation}
\frac{n_e}{n_1} = \frac{\erfc(\nu_1)}{2} , \label{eq:ntrun}
\end{equation}
\begin{equation}
\frac{\vc{V}_e}{v_{T1}} = \frac{n_1}{n_e} \frac{ e^{-\nu_1^2}}{2 \sqrt{\pi} } \hat{x} , \label{eq:vtrun}
\end{equation}
and
\begin{equation}
\frac{T_e}{T_1} =  1 + \frac{2 \nu_1}{3} \frac{V_e}{v_{T1}} - \frac{2}{3} \frac{V_e^2}{v_{T1}^2} \label{eq:ttrun}
\end{equation}
for the electron density, flow velocity and temperature, respectively. These are shown in Fig.~\ref{fg:moments} as a function of the cutoff parameter $\nu_1 = v_c/v_{T1}$. The coordinates are aligned so that $\nu_1 \leq 0$. 

We consider only waves that propagate perpendicular to the boundary, such that $\vc{k} = k \hat{x}$. Taking the model distribution from Eq.~(\ref{eq:ftrun}) for electrons, and a flowing Maxwellian distribution for the ions, the dielectric function from Eq.~(\ref{eq:dielec}) reduces to
\begin{equation}
\hat{\varepsilon} = 1 - \frac{\omega_{pi}^2}{k^2 v_{Ti}^2} Z^\prime \biggl( \frac{\omega - \vc{k} \cdot \vc{V}_i}{k v_{Ti}} \biggr) - \frac{\omega_{p1}^2}{k^2 v_{T1}^2} Z^\prime \biggl( \nu_1 , \frac{\omega}{k v_{T1}} \biggr) . \label{eq:eptrun}
\end{equation}
Next, numerical solutions and asymptotic approximations will be used to study the dispersion relations of ion-acoustic and Langmuir frequency waves determined from roots of Eq.~(\ref{eq:eptrun}). We assume the flow is only in the parallel direction, $\vc{V}_i = V_i \hat{x}$, and the abbreviated notation $\vc{k} \cdot \vc{V}_i = k V_i$ is used. 

\subsection{Ion-acoustic waves\label{sec:iatrun}}

An approximate analytic dispersion relation for ion-acoustic waves is obtained from the large argument expansion of the plasma dispersion function for the ion term $(\omega - \vc{k} \cdot \vc{V}_i)/(kv_{Ti}) \gg 1$, and the small argument expansion of the incomplete plasma dispersion function for the electron term $\omega /(kv_{T1}) \ll 1$ using Eqs.~(\ref{eq:ode}) and (\ref{eq:sarg}). Applying these asymptotic expansions to Eq.~(\ref{eq:eptrun}), the dielectric function reduces to
\begin{equation}
\hat{\varepsilon} = 1 + \frac{\alpha}{k^2 \lambda_{D1}^2} - \frac{\omega_{pi}^2}{(\omega - k V_i)^2} + \frac{i \delta w}{k^2 \lambda_{D1}^2} \label{eq:epia}
\end{equation}
in which $w = \omega/(kv_{T1})$, 
\begin{equation}
\alpha = \frac{1}{2} \biggl[ \erfc (\nu_1) - \frac{e^{-\nu_1^2}}{\sqrt{\pi} \nu_1} \biggr]  ,
\end{equation}
and 
\begin{equation}
\delta = \sqrt{\pi} H(w_R - \nu_1) - \frac{i}{2} \biggl( \frac{E_1 (\nu^2)}{\sqrt{\pi}} + \frac{e^{-\nu^2}}{\sqrt{\pi} \nu^2} \biggr) .
\end{equation}

\begin{figure}
\begin{center}
\includegraphics{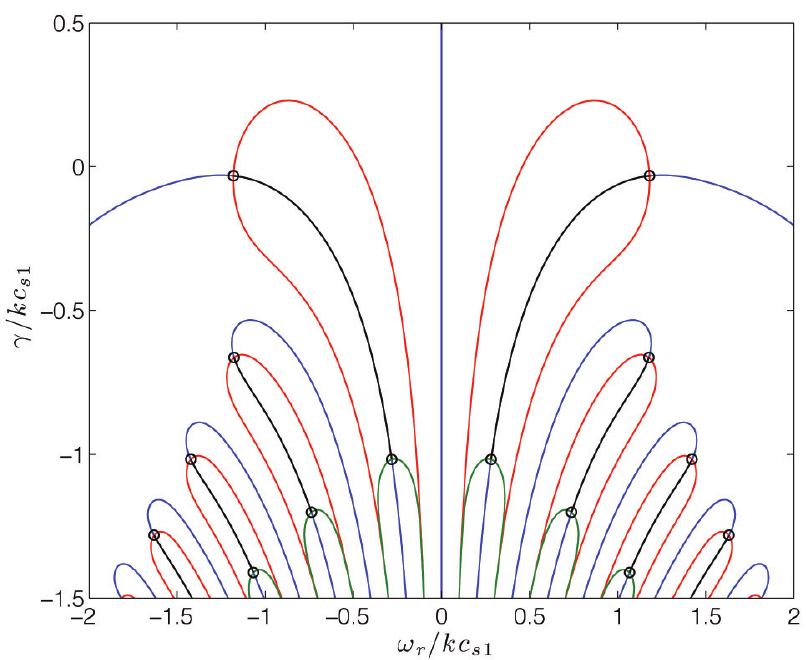}
\caption{Cuts of the intersection of the real and imaginary parts of the electrostatic dielectric function with the $[0,0]$ plane. The real part is shown for $k=10^{-2}$ (red lines) and $k=10^{2}$ (green lines). The imaginary part is the same for each $k$ (blue lines). The black lines show the dispersion relation of various modes as $k$ varies from $10^{-2}$ to $10^2$. }
\label{fg:ia_dielec}
\end{center}
\end{figure}

\begin{figure}
\begin{center}
\includegraphics{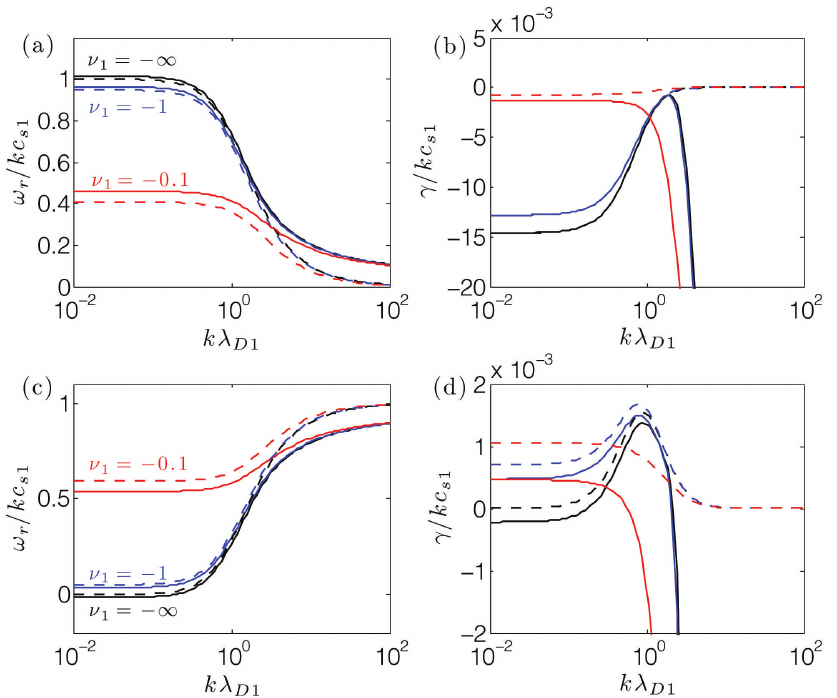}
\caption{The ion-acoustic dispersion relation computed numerically from Eq.~(\ref{eq:eptrun}) (solid lines) and from the approximate analytic formula from Eqs.~(\ref{eq:iatrun}) and (\ref{eq:iatrun2}) (dashed lines) for three values of the cutoff parameter $\nu_1 = -\infty, -1$ and $-0.1$. The parameters $T_1/T_i =100$, $Z_i=1$, and $m_i/m_e =1836$ were used in each case. Panels (a) and (b) show the real and imaginary components for no flow $V_i=0$, while (c) and (d) correspond to sonic flow $V_i = c_{s1}$. }
\label{fg:ia_disp}
\end{center}
\end{figure}

Assuming that $\delta w \ll 1 + k^2 \lambda_{D1}^2$, and that $\omega_R \gg \gamma$, the dispersion relation obtained from the roots of Eq.~(\ref{eq:epia}) is
\begin{equation}
\omega_\pm = \omega_{R,\pm} \biggl[ 1 \mp i Z_i \sqrt{\frac{n_e}{n_1}} \sqrt{\frac{\pi m_e}{8M_i}} \frac{H(w_R - \nu_1)}{(\alpha + k^2 \lambda_{D1}^2)^{3/2}} \biggr] \label{eq:iatrun}
\end{equation}
in which
\begin{equation}
\omega_{R,\pm} = k V_i \pm \sqrt{\frac{n_e}{n_1}} \frac{Z_i k c_{s1}}{\sqrt{\alpha + k^2 \lambda_{D1}^2}} \label{eq:iatrun2}
\end{equation} 
is the real frequency of the waves. The notation in Eqs.~(\ref{eq:iatrun}) and (\ref{eq:iatrun2}) assumes either the top or bottom set of $\pm$ symbols is used (not any permutation). The conventional ion-acoustic dispersion relation is returned in the limit $n_e/n_1 \rightarrow 1$, $\alpha \rightarrow 1$, and $H(w_R - \nu) \rightarrow 1$. 

Numerical dispersion relations have been obtained directly from Eq.~(\ref{eq:eptrun}). For ion-acoustic frequency waves, it is convenient to write Eq.~(\ref{eq:eptrun}) in terms of the dimensionless frequency and growth rate $\omega_r/kc_{s1}$ and $\gamma/kc_{s1}$. Six free parameters then remain: $k \lambda_{D1},\ \nu_1,\ T_1/T_i,\ V_i/c_{s1},\ m_e/m_i$ and $Z_i$. The dispersion relation for various roots at fixed $k \lambda_{D1}$ are obtained from the points where both the real and imaginary parts of the dielectric function, Eq.~(\ref{eq:eptrun}), vanish. Figure~\ref{fg:ia_dielec} shows cuts of each component in the $[0, 0]$ plane. Here the red and blue lines denote the real and imaginary components for $k\lambda_{D1} = 10^{-2}$. The other free parameters chosen in this plot are $\nu_1 = -\infty ,\ T_1/T_i =10,\ V_i/c_{s1} = 0,\ m_i/m_e =1836$ and $Z_i=1$. The upper set of circles denote the roots for this wavenumber, which each correspond to a different wave. Dispersion relations are obtained by stepping $k\lambda_{D1}$ and determining the root at each step. This procedure maps the black lines in Fig.~\ref{fg:ia_dielec} as $k\lambda_{D1}$ varies from $10^{-2}$ to $10^2$. The green lines show the cut of the real component of $\hat{\varepsilon}$ for $k\lambda_{D1} = 10^2$ and the imaginary component is essentially unchanged (blue lines). Figure~\ref{fg:ia_dielec} shows that these ion-acoustic dispersion relations lie along the zero-plane cuts of the imaginary component of $\hat{\varepsilon}$. 

Figure~\ref{fg:ia_disp} shows example dispersion relations obtained with this method for three values of the cutoff parameter: $\nu_1 = v_c/v_{T1} = -\infty ,\ -1$ and $-0.1$. Here the frequencies are normalized to $k c_{s1}$. Panels (a) and (b) take $V_i=0$ and look at the least damped root with $\omega_r$ positive, while (c) and (d) take $V_i = c_{s1}$ and look at the root where $\omega_r/kc_{s1}$ is small, which is unstable for a range of $k \lambda_{D1}$ (there is also a higher frequency mode with $\omega_r \sim 2 kc_{s1}$ in this case, which is damped).  The other parameters used are $T_1/T_i = 100,\ m_i/m_e =1836$ and $Z_i =1$. Solid lines represent the numerical solutions and dashed lines the analytic approximations from Eqs.~(\ref{eq:iatrun}) and (\ref{eq:iatrun2}). The curves for $\nu_1 = -\infty$ correspond to the conventional ion-acoustic wave in a Maxwellian plasma. The figure shows that this is not significantly modified for $\nu_1$ as large as $-1$. Thus, modifications to the ion-acoustic dispersion relation due to the depleted electron interval near floating surfaces is negligible. For floating surfaces, $\nu_1$ is typically a few times $-1$. Significant ($\sim 50 \% $) corrections arise in the $\nu_1 = -0.1$ case. Such a case is relevant near isolated boundaries, such as probes, biased so that the potential drop from the plasma to probe is less than the floating potential. Boundaries biased more positive than the plasma often result in electron distributions with $\nu_1 =0$. 

Figure~\ref{fg:ia_disp} also shows that the analytic approximations from Eqs.~(\ref{eq:iatrun}) and (\ref{eq:iatrun2}) are typically within the expected  $\mathcal{O}(\sqrt{T_i/T_1}) \sim 10 \%$ accuracy for $k \lambda_{D1} \lesssim 1$. One exception is the growth rate for the $\nu_1 = -0.1$ case in panel (d), which is off by $\sim 50 \%$ for small $k\lambda_{D1}$. One interesting feature of the depleted interval is that it can have a greater affect on the growth rate than it does the real frequency; as shown in panels (c) and (d). For $\nu_1 = -1$, depletion has a negligible affect on the real frequency, but destabilizes low-$k$ waves. However, the depletion produces competing effects, as demonstrated by the result that increasing $\nu_1$ further to $-0.1$ enhances the stabilizing effect of finite $k \lambda_{D1}$. This may be because the electron-ion temperature ratio, see Eq.~(\ref{eq:ttrun}), is substantially reduced in this case. 

\subsection{Langmuir waves\label{sec:langt}} 

To obtain an approximate analytic dispersion relation for Langmuir waves ($\omega \sim \omega_{pe}$), the ion term in Eq.~(\ref{eq:eptrun}) is $\mathcal{O}(m_e/m_i)$ and will be neglected. Taking the large argument expansion of the electron term from Eq.~(\ref{eq:larg}), $Z^\prime \simeq [\erfc(\nu)/2 + \delta (\bar{w})]/w^2$, where $\delta$ keeps terms up to fourth order terms, and $\bar{w}$ uses the lowest order solution of $\omega$ ($\bar{w} = \omega_{pe}/kv_{T1}$). Using these in Eq.~(\ref{eq:eptrun}) provides the dispersion relation
\begin{equation}
\omega^2 \simeq \omega_{pe}^2 + \frac{3}{2} k^2 v_{T1}^2 + 2kV_e \biggl( \omega_{pe} + \frac{3}{2} k v_{T1} \nu_1 \biggr)  , \label{eq:ltrun}
\end{equation}
in which $\omega_{pe}^2 = \erfc(\nu) \omega_{p1}^2/2$ is the plasma frequency based on the density moment and $V_e$ is the electron flow moment speed from Eq.~(\ref{eq:vtrun}). 

\begin{figure}
\begin{center}
\includegraphics{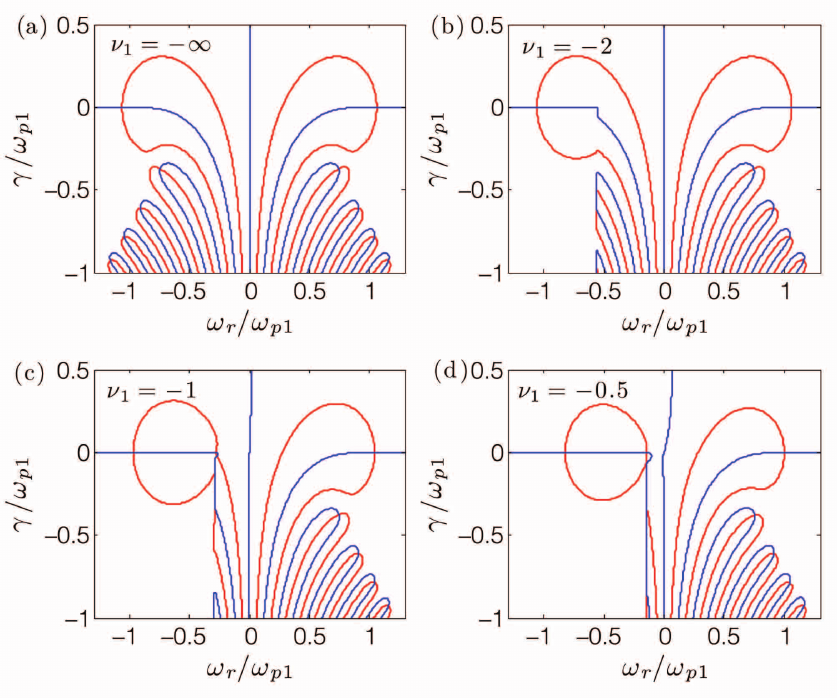}
\caption{Contours of $\Re \lbrace \hat{\varepsilon} \rbrace = 0$ (red) and $\Im \lbrace \hat{\varepsilon} \rbrace =0$ (blue) in the complex frequency plane ($\omega = \omega_R + i \gamma$) for Langmuir-frequency waves from Eq.~(\ref{eq:eptlang}) taking $k\lambda_{D1} = 0.2$.}
\label{fg:lang_dielec}
\end{center}
\end{figure}

\begin{figure}
\begin{center}
\includegraphics{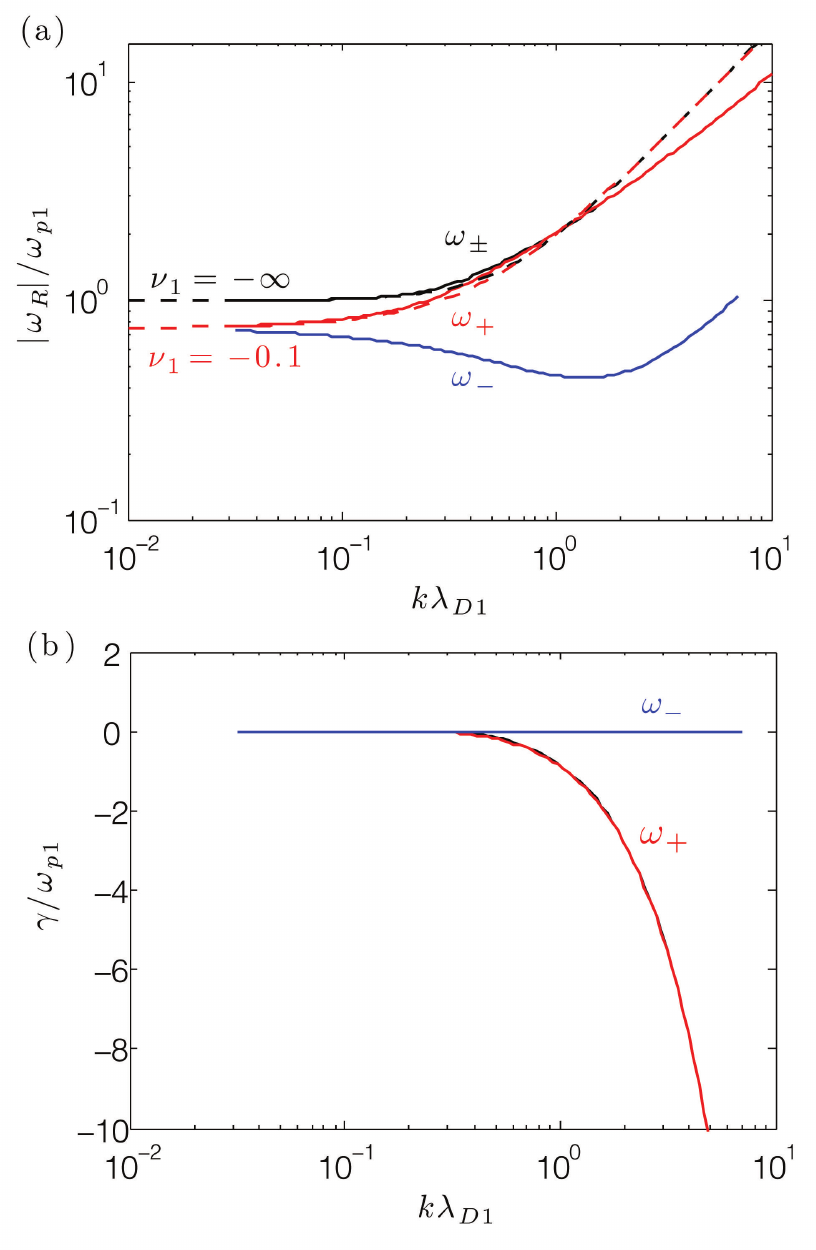}
\caption{Langmuir wave dispersion relation calculated numerically from Eq.~(\ref{eq:eptlang}) (solid lines) and analytically from Eq.~(\ref{eq:ltrun}) (dashed lines). (a) shows the real frequency for positive phase velocity ($\omega_+$) and negative phase velocity ($\omega_-$) modes. (b) shows the growth rate for each mode. The damping rate for $\omega_{\pm} (\nu_1 = -\infty)$ and $\omega_+ (\nu_1 = -0.1)$ lie on top of one another. }
\label{fg:lang_t}
\end{center}
\end{figure}

The ion term in Eq.~(\ref{eq:eptrun}) is also neglected for the numerical analysis of Langmuir waves, which are determined from the roots of
\begin{equation}
\hat{\varepsilon} = 1 - \frac{\omega_{p1}^2}{k^2 v_{T1}^2} Z^\prime \biggl( \nu_1 , \frac{\omega}{k v_{T1}} \biggr) . \label{eq:eptlang}
\end{equation}
Figure~\ref{fg:lang_dielec} shows some interesting features of the zeros of Eq.~(\ref{eq:eptlang}) as the cutoff parameter increases from $-\infty$. In these plots, $k \lambda_{D1} = 0.2$ was chosen. In the complete case ($\nu_1 = -\infty$) the two usual Langmuir roots are found $\omega = \pm \omega_{p1}$. The dispersion of each of these is identical, except for the sign of the phase speed. For $\nu_1 = -2$ (at this $k\lambda_{D1}$), these two weakly-damped waves persist nearly unaffected by the depleted region in velocity-space. However, several damped modes are strongly affected by the depletion, leading to an asymmetry in $\pm \omega_R$ that is not present in the complete case. As $\nu_1$ is increased further to $-1$, this asymmetry extends to the usual Langmuir roots as well. Here $|\omega_{R-}| < |\omega_{R+}|$, but only slightly so for $\nu_1 =-1$. A more substantial asymmetry in the real frequency of the two oppositely directed modes is found as the cutoff parameter increases further. 

Figure \ref{fg:lang_t} shows the Langmuir wave dispersion relation for each of these modes at two values of the cutoff parameter: $\nu_1 = -\infty$ and $-0.1$. The complete case is symmetric, so $\omega_{R+} = - \omega_{R-}$, as is typical. As the cutoff parameter increases two effects set in: (1) there is a shift in the frequency at asymptotically low wavenumber, and (2) the two oppositely directed modes no longer have the same absolute frequency. Effect (1) is simply due to the $\omega_{p1}$ normalization. Equation~(\ref{eq:ltrun}) shows that $\omega \simeq \pm \omega_{pe}$ at asymptotically small wavenumbers, where $\omega_{pe}$ is the electron plasma frequency based on the total electron density. For $\nu_1 =-0.1$, $\omega_{pe}/\omega_{p1} =\sqrt{n_e/n_1} \approx 1/\sqrt{2}$, which accounts for this shift. Effect (2) sets in at higher wavenumbers, which the asymptotic theory does not capture. Here, the magnitude of the real frequency can be signficantly smaller for the mode with phase velocity in the depleted region, in comparison to the oppositely directed mode. Because the electron density vanishes in this region of velocity phase-space, their are no resonant electrons, and hence no Landau damping. This mode is undamped, even for large $k \lambda_{D1}$, in contrast to the strong damping that affects the mode with positive phase velocity. This can be seen in Fig.~\ref{fg:lang_t}b.

\section{Waves in the presence of a partially depleted electron distribution\label{sec:pvsd}} 

Scattering and ionization cause the passing interval of the electron distribution to be refilled as a function of distance from the boundary. Here, the discussion of the last section is generalized to include a passing population (labeled population $2$), characterized by $T_2 \leq T_1$. The electron distribution then takes the form
\begin{eqnarray}
f_e = \frac{e^{-v_\perp^2/v_{T1}^2}}{\pi^{3/2}v_{T1}^2} \label{eq:fdep1}
\left\lbrace \begin{array}{cc}
n_2 \exp(-v_x^2/v_{T2}^2)/v_{T2}, & v_x < v_c \\
n_1 \exp(-v_x^2/v_{T1}^2)/v_{T1},  & v_x \geq v_c
\end{array} \right. 
\end{eqnarray}
where $v_c \leq 0$. For absorbing boundaries $f_e$ is expected to be continuous, in which case 
\begin{equation}
n_2 = n_1 \sqrt{\frac{T_2}{T_1}} e^{\nu_2^2 - \nu_1^2} . \label{eq:n2}
\end{equation}
Although the continuous case of Eq.~(\ref{eq:n2}) is considered here, the results can be generalized in a straight-forward way to electron-emitting boundaries by applying a model for $n_2$ that accounts for this additional population of electrons in the passing interval. 

Applying Eqs.~(\ref{eq:fdep1}) and (\ref{eq:n2}) to the moment definitions of the fluid variables provides
\begin{equation}
\frac{n_e}{n_1} = \frac{\erfc (\nu_1)}{2} +  \frac{n_2}{n_1} \frac{\erfc (- \nu_2)}{2}  , \label{eq:dens2}
\end{equation}
\begin{equation}
\frac{\vc{V}_e}{v_{T1}} = \frac{n_1}{n_e} \frac{ (1 - T_2/T_1)}{2 \sqrt{\pi}} e^{-\nu_1^2} \hat{x} ,  \label{eq:vel2}
\end{equation}
and
\begin{equation}
\frac{T_e}{T_1} = \frac{2}{3} \biggl[ 1 + \frac{1}{2} \frac{T_2}{T_1} + \frac{V_e}{v_{T1}} \biggl( \nu_1 - \frac{V_e}{v_{T1}} + \frac{\sqrt{\pi}}{2} \erfc(\nu_1) e^{\nu_1^2} \biggr) \biggr]
\end{equation}
for the electron density, flow velocity and temperature, respectively. These are shown in Fig.~\ref{fg:moments} as a function of the cutoff parameter $\nu_1 = v_c/v_{T1}$ (note $\nu_2 = \nu_1 \sqrt{T_1/T_2}$) for a few values of the temperature ratio $T_2/T_1$. Putting Eq.~(\ref{eq:fdep1}) into (\ref{eq:dielec}), the linear plasma dielectric takes the form 
\begin{align}
\hat{\varepsilon} &= 1 - \frac{\omega_{pi}^2}{k^2 v_{Ti}^2} Z^\prime \biggl( \frac{\omega - \vc{k} \cdot \vc{V}_i}{k v_{Ti}} \biggr)  \label{eq:ep2temp} \\ \nonumber
&- \frac{ \omega_{p1}^2}{k^2 v_{T1}^2} \biggl\lbrace Z^\prime (\nu_1, w_1) + \frac{T_1 n_2}{T_2 n_1} [ Z^\prime (w_2) - Z^\prime (\nu_2 , w_2)] \biggr\rbrace  . 
\end{align}
Next, the ion-acoustic and Langmuir waves from the previous section are revisited using Eq.~(\ref{eq:ep2temp}), which includes a non-vanishing tail population. 

\subsection{Ion-acoustic waves}

To obtain an approximate analytic dispersion relation, we apply to Eq.~(\ref{eq:ep2temp}) the usual ordering: $(\omega - \vc{k} \cdot \vc{V}_i)/kv_{Ti} \gg1$, $\omega/kv_{T1} \ll 1$ and $\omega / kv_{T2} \ll 1$. However, we note at the outset that the approximate dispersion relation obtained in this manner cannot be expected to reduce to Eq.~(\ref{eq:iatrun}) in the $T_2 \rightarrow 0$ limit because this would violate the $\omega/ kv_{T2} \ll 1$ assumption. Estimating $\omega/kv_{T2} \sim c_{s1}/v_{T2}$, the $\omega/kv_{T2} \gg 1$ ordering requires $T_2 \gg T_1 m_e/m_i$. Following the same procedure as Sec.~\ref{sec:iatrun} provides the approximate ion-acoustic wave dispersion relation
\begin{equation}
\omega_\pm = \omega_{R,\pm} \biggl[1 \mp i Z_i \sqrt{\frac{n_e}{n_1}} \sqrt{\frac{\pi m_e}{8 M_i}} \frac{\Theta (\omega_R)}{(\bar{\alpha} + k^2 \lambda_{D1}^2)^{3/2}} \biggr] \label{eq:iadepl}
\end{equation}
in which 
\begin{equation}
\omega_{R, \pm} = k V_i \pm \sqrt{\frac{n_e}{n_1}} \frac{Z_i k c_{s1}}{\sqrt{\bar{\alpha} + k^2 \lambda_{D1}^2}} ,
\end{equation}
is the real frequency. Here
\begin{equation}
\bar{\alpha} = \frac{\erfc (\nu_1)}{2} + \frac{T_1 n_2}{T_2 n_1} \frac{\erfc (- \nu_2)}{2} 
\end{equation}
and
\begin{equation}
\Theta = H(w_{R1} - \nu_1) + H(\nu_2 - w_{R2}) \biggl( \frac{T_1}{T_2} \biggr)^{3/2} \frac{n_2}{n_1} e^{-w_{R2}^2}  .
\end{equation} 
The conventional ion-acoustic dispersion relation can be returned in the $T_2/T_1 \rightarrow 1$ limit (in which case $n_e/n_1,\ \bar{\alpha}$ and $\Theta \rightarrow 1$). 

\begin{figure}
\begin{center}
\includegraphics{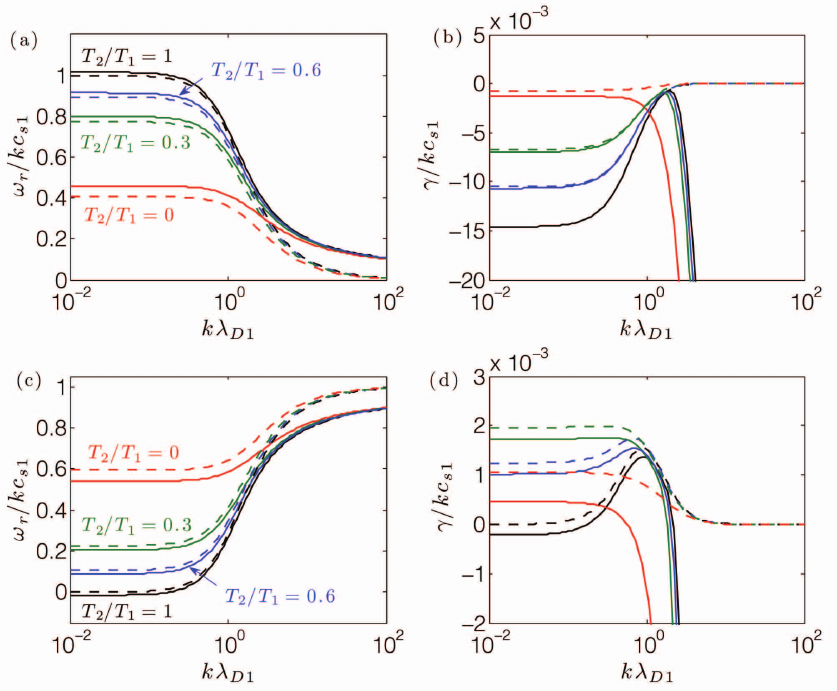}
\caption{The ion-acoustic dispersion relation computed numerically from Eq.~(\ref{eq:ep2temp}) (solid lines) and from the approximate analytic formula (dashed lines) for values of the temperature ratio: $T_2/T_1 = 1,\ 0.6,\ 0.3$ and $0$. The parameters $\nu_1 = -0.1$, $T_1/T_i =100$, $Z_i=1$, and $m_i/m_e =1836$ were used in each case. Panels (a) and (b) show the real and imaginary components for no flow $V_i=0$, while (c) and (d) correspond to sonic flow $V_i = c_{s1}$.}
\label{fg:ia_depl}
\end{center}
\end{figure}

Numerical solutions of the dispersion relation were obtained from Eq.~(\ref{eq:ep2temp}) in terms of $\omega/kc_{s1}$ using the same method described in Sec.~\ref{sec:iatrun}. Here seven free parameters must be set: $k\lambda_{D1},\ \nu_1,\ T_2/T_1,\ T_1/T_i,\ V_i/c_{s1},\ m_e/m_i$ and $Z_i$. Figure~\ref{fg:ia_depl} shows the real frequency and growth rate as a function of $k \lambda_{D1}$ for a few trapped-passing temperature ratios for stationary ($V_i = 0$) and sonically flowing ($V_i = c_{s1}$) ion distributions. The other free parameters chosen were: $\nu_1 = -0.1,\ T_1/T_i=100,\ m_i/m_e = 1836$ and $Z_i =1$. Figure~\ref{fg:ia_depl} shows a transition between the conventional ion-acoustic wave ($T_2/T_1 = 1$) and the fully depleted wave ($T_2/T_1 = 0$) as $T_2/T_1$ decreases. It also shows that the approximate formula from Eq.~(\ref{eq:iadepl}) typically captures the small $k \lambda_{D1}$ behavior of the real frequency and growth rate to within the expected $\mathcal{O}(\sqrt{T_e/T_i}) \sim 10\%$ accuracy. Equation~(\ref{eq:iatrun}) was used for the $T_2/T_1 = 0$ case, since Eq.~(\ref{eq:iadepl}) is not expected to hold in this limit. 

\subsection{Langmuir waves} 

An approximate analytic expression for Langmuir waves can be obtained from Eq.~(\ref{eq:ep2temp}) using the same procedure as Sec.~\ref{sec:langt}. Neglecting the ion term, and obtaining up to fourth order in the large argument expansion of the electron terms, this procedure provides the dispersion relation 
\begin{align}
& \omega^2 \simeq \omega_{pe}^2 + \frac{3}{2} k^2 v_{T1}^2 \label{eq:langdepl} \\ \nonumber
&+ 2 k V_e \biggl[ \omega_{pe} + \frac{3}{2} k v_{T1} \biggl( \nu_1 - \sqrt{\frac{T_2}{T_1}} \frac{\erfc (-\nu_2)}{2} e^{\nu_2^2} \biggr) \biggr] ,
\end{align}
in which $\omega_{pe}$ is the electron plasma frequency based on the total electron density $n_e$ from Eq.~(\ref{eq:dens2}) and $V_e$ is the flow-moment speed from Eq.~(\ref{eq:vel2}). Equation~(\ref{eq:langdepl}) reduces to the conventional Langmuir wave dispersion relation in the limit $T_2/T_1 \rightarrow 1$ ($V_e \rightarrow 0$) and to Eq.~(\ref{eq:ltrun}) in the limit $T_2 \rightarrow 0$. 

\begin{figure}
\begin{center}
\includegraphics{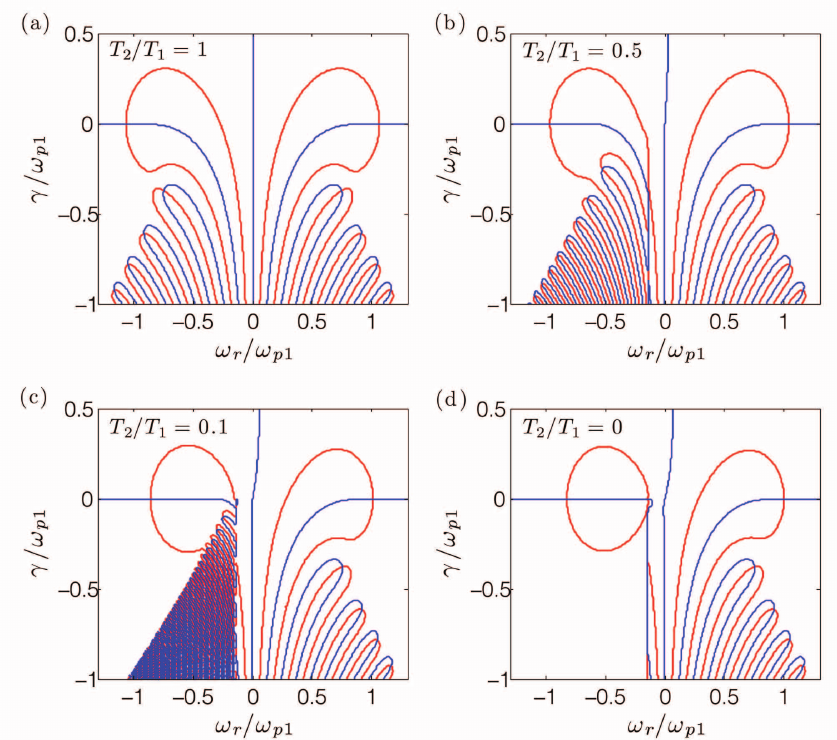}
\caption{Contours of $\Re \lbrace \hat{\varepsilon} \rbrace = 0$ (red) and $\Im \lbrace \hat{\varepsilon} \rbrace =0$ (blue) in the complex frequency plane ($\omega = \omega_R + i \gamma$) for Langmuir-frequency waves from Eq.~(\ref{eq:epdlang}) taking $k\lambda_{D1} = 0.2$ and $\nu_1 = -0.5$. Panel (d) solves Eq.~(\ref{eq:eptlang}) for the $T_2 =0$ case. }
\label{fg:l_dielec_depl}
\end{center}
\end{figure}

\begin{figure}
\begin{center}
\includegraphics{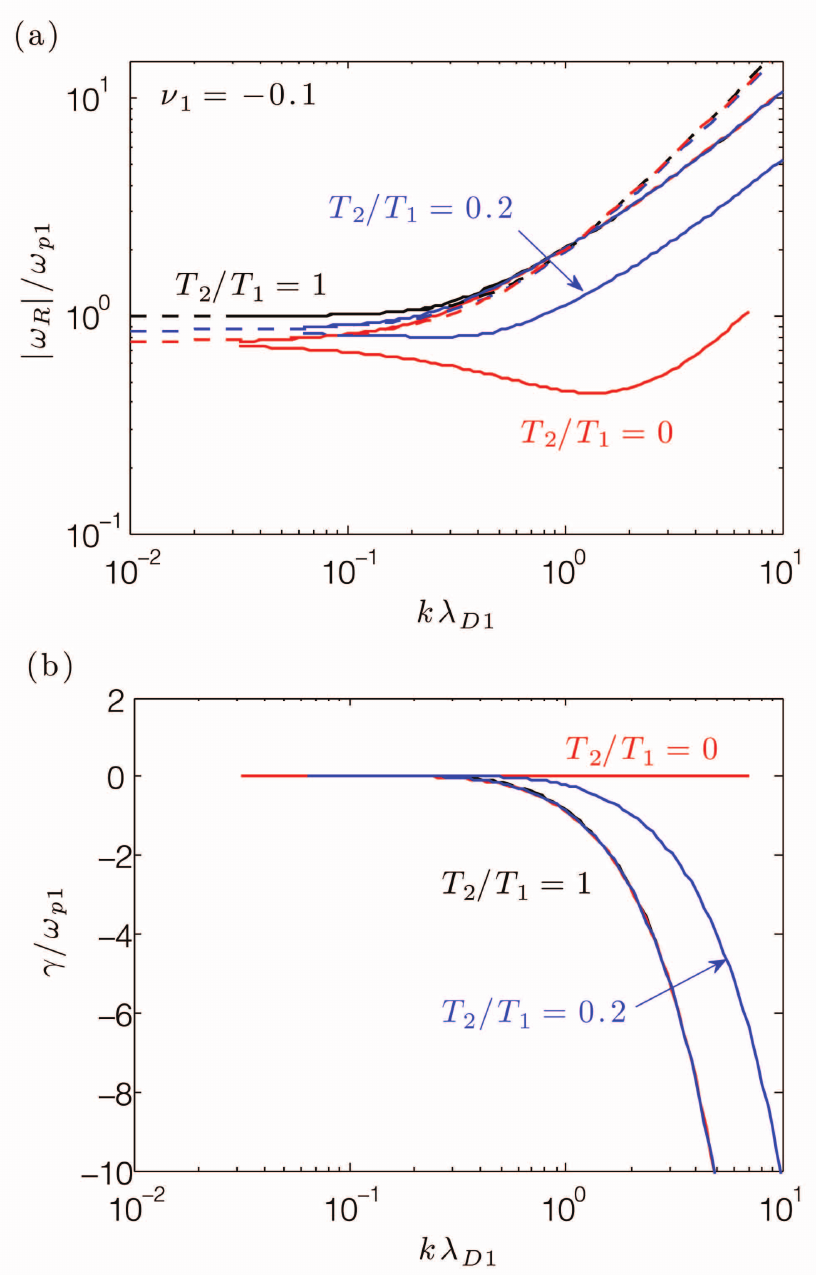}
\caption{Langmuir wave dispersion relation calculated numerically from Eq.~(\ref{eq:epdlang}) (solid lines) and analytically from Eq.~(\ref{eq:langdepl}) (dashed lines) for $\nu_1 = -0.1$ and three values of the temperature ratio: $T_2/T_1 = 1,\, 0.2$ and $0$. (a) shows the real frequency for negative phase velocity modes, and (b) shows the growth rate for each mode. }
\label{fg:lang_d}
\end{center}
\end{figure}

As in Sec.~\ref{sec:langt}, the ion term is dropped for the numerical analysis for the high-frequency Langmuir waves, so Eq.~(\ref{eq:ep2temp}) reduces to
\begin{equation}
\hat{\varepsilon} = 1 - \frac{ \omega_{p1}^2}{k^2 v_{T1}^2} \biggl\lbrace Z^\prime (\nu_1, w_1) + \frac{T_1 n_2}{T_2 n_1} [ Z^\prime (w_2) - Z^\prime (\nu_2 , w_2)] \biggr\rbrace  . \label{eq:epdlang}
\end{equation}
Figure~\ref{fg:l_dielec_depl} shows how the zeros of $\Re \lbrace \hat{\varepsilon} \rbrace$ and $\Im \lbrace \hat{\varepsilon} \rbrace$, and the corresponding dispersion relations, change as the trapped-passing temperature ratio is varied. The parameters $\nu_1 = -0.5$ and $k \lambda_{D1} = 0.2$ have been chosen for this figure. The usual symmetric roots are obtained for the complete case ($T_2/T_1 = 1$). As the temperature ratio drops, this symmetry is broken and the density of damped roots increases substantially for $\omega_R/\omega_{p1} \lesssim \nu_1 k \lambda_{D1} \approx -0.1$. As the temperature ratio drops, the asymmetry between the oppositely directed modes increases, and the Landau damping decreases. 

Figure \ref{fg:lang_d} shows the Langmuir wave dispersion relation for $\nu_1 = -0.1$ and three values of the temperature ratio $T_2/T_1 =1,\ 0.2$ and $0$. The general behavior is similar to the totally depleted case from Fig.~\ref{fg:lang_t}. The $T_2/T_1 = 0.2$ line shows a transition between the two limiting cases. Figure~\ref{fg:lang_d}b shows that the undamped modes extend to higher wavenumbers for cold, but nonzero, tail temperature. However, Landau damping does not vanish completely for finite tail temperature. The approximate formula from Eq.~(\ref{eq:langdepl}) accurately predict the wave frequency at small wavenumber. 

\section{Summary\label{eq:sum}}

Plasmas, especially in the presence of potential barriers, can be far from equilibrium. Nevertheless, models can often be constructed to describe the distribution functions using Maxwellians in finite velocity-space intervals, where each interval is characterized by different effective densities, temperatures and flow speeds. Whenever such models are applied to calculate the linear plasma dielectric response, the incomplete plasma dispersion function, defined by Eq.~(\ref{eq:ipdf}), will arise. Several properties of this function, such as asymptotic expansions and other approximations, that are useful for applying it to calculate wave dispersion relations were reviewed in Sec.~\ref{sec:prop}. In Secs.~\ref{sec:vsd} and \ref{sec:pvsd}, this function was used to develop quantitative dispersion relations for ion-acoustic and Langmuir waves in plasmas near absorbing boundaries. The passing interval of the electron distribution is depleted in density compared to the trapped interval in this circumstance. Substantial modifications to the dispersion relations were shown to arise if the passing interval is sufficiently cold, and the trapped-passing boundary is not too far onto the tail of the distribution. These conditions are particularly interesting near probes biased near (or above) the plasma potential. 

\appendix
\section{Derivation of generating function for $Z_n$ \label{sec:zngen}}
\setcounter{section}{1}

An outline of the derivation of Eq.~(\ref{eq:zngen}), which relates Eq.~(\ref{eq:zgen}) to derivatives of $Z(\nu, w)$, is provided in this appendix. First, it is useful to apply $n$ derivatives to Eq.~(\ref{eq:ipdf}) to give
\begin{equation}
Z^{(n)} (\nu, w) = \frac{n!}{\sqrt{\pi}} \int_\nu^\infty dt \frac{e^{-t^2}}{(t-w)^{n+1}} . \label{eq:nder}
\end{equation}
Applying
\begin{align}
\frac{e^{-t^2}}{(t - w)^{n+1}} &= \frac{1}{n!} \frac{ (d^n/dt^n) e^{-t^2}}{t - w} \\ \nonumber
&- \frac{d}{dt} \biggl[ \sum_{m=0}^{n-1} \frac{(n - m - 1)!}{n!} \frac{ (d^m/dt^m) e^{-t^2}}{(t - w)^{n-m}} \biggr] 
\end{align}
and the relation
\begin{equation}
(d^n/dt^n) e^{-t^2} = (-1)^n e^{-t^2} H_n (t) ,
\end{equation}
Eq.~(\ref{eq:nder}) can be rearranged to show
\begin{align}
& \frac{1}{\sqrt{\pi}} \int_\nu^\infty dt \frac{H_n(t) e^{-t^2}}{t - w} = (-1)^n Z^{(n)} \label{eq:hint} \\ \nonumber
&- \frac{(-1)^n}{\sqrt{\pi}} \sum_{m=1}^{n-1} \frac{(n - m - 1)! (-1)^m H_m (\nu) e^{-\nu^2}}{(\nu - w)^{n-m}} .
\end{align}

Using the identity 
\begin{equation}
t^n = \frac{1}{2^n} \sum_{l=0}^n d_l (n) H_l (t)
\end{equation}
in Eq.~(\ref{eq:zgen}) shows 
\begin{equation}
Z_n (\nu , w) = \frac{1}{2^n} \sum_{l=0}^n d_l (n) \frac{1}{\sqrt{\pi}} \int_\nu^\infty dt \frac{H_l (t) e^{-t^2}}{t - w} . \label{eq:znmid}
\end{equation}
Finally, putting Eq.~(\ref{eq:hint}) into Eq.~(\ref{eq:znmid}) completes the derivation of Eq.~(\ref{eq:zngen}). 

\begin{acknowledgments}

The author thanks Prof.\ F.\ Skiff for reading and commenting on the manuscript and Prof.\ C.\ C.\ Hegna for helpful discussions. This research was supported in part by an appointment to the U.S. Department of Energy Fusion Energy Postdoctoral Research Program administered by the Oak Ridge Institute for Science and Education, and in part under the auspices of the National Nuclear Security Administration of the U.S. Department of Energy at Los Alamos National Laboratory under Contract No. DE-AC52-06NA25396.


\end{acknowledgments}


\end{document}